\documentclass[a4paper]{article}
\usepackage{graphicx}
\begin{document}

\centerline{\bf\Large Spin and parity of a possible baryon antidecuplet}
\vspace{4ex}
\centerline{C. Dullemond}
\centerline{Institute for Theoretical Physics}
\centerline{University of Nijmegen}
\centerline{Nijmegen, The Netherlands}

\section*{Abstract}

The recently postulated existence of a baryon antidecuplet [1] can be
reproduced in strong coupling theory in which a bare baryon spin $1/2$ octet
interacts with an octet of pseudoscalar mesons. When a suitable mixture of
F- and D-type Yukawa couplings is chosen the dressed baryons group
themselves into an infinite number of SU(3) multiplets of which the ground
state turns out to be a spin $1/2$ baryon octet and the first excited state
is a spin $3/2$ baryon decuplet. Then follows a spin $1/2$ baryon
antidecuplet. All states of the spectrum have positive parity. If the
hypothetical baryon antidecuplet can be identified with the antidecuplet in
the strong coupling spectrum then a positive parity and a spin $1/2$ is
predicted for this multiplet.

\section{Introduction}

Recently new baryon resonances have been found which are ``exotic'' in the
sense that they cannot be considered as three-quark states [1]. As
``pentaquark baryons'' they may fit into a baryon antidecuplet. It is the
purpose of this Letter to remind of an earlier attempt to generate baryon
spectra, namely strong coupling theory. Nonrelativistic strong coupling
theory has played an important role in the early days of field theory and
nuclear physics [2]. With the advent of flavor SU(3) particle multiplets, it
has been tried to explain the existence of the spin $1/2$ baryon octet and
the spin $3/2$ baryon decuplet from field theory models with strong coupling
between a ``bare'' baryon octet and a meson octet. Although the physical ideas
underlying the model are presently unacceptable, the obtained baryon
spectrum was in remarkable agreement with experimental data at that time and
may still be of some relevance. Results were first given for a spinless
baryon octet in interaction with a scalar meson octet [3], later followed
the results for a spin $1/2$ baryon octet interacting with a pseudoscalar
meson octet [4,5]. The interactions in the model are of Yukawa type and are
a mixture of F- and D-type coupling such that breaking of
multiplet-antimultiplet symmetry in the final spectrum is guaranteed. While
in the spinless case the spectrum is a continuous function of the F/D ratio,
in the more realistic case of spin $1/2$ baryons in interaction with
pseudoscalar mesons the spectrum turns out to be much more rigid.

\section{The method}
Starting point is the following hamiltonian:
\begin{equation}
H = \frac{1}{2}\left(p^{\alpha i}_{\beta j} p^{\beta j}_{\alpha i} 
+ \mu^2 q^{\alpha i}_{\beta j} q^{\beta j}_{\alpha i}\right) 
+ g_1 \bar B^{\gamma}_{\delta i} q^{\delta i}_{\varepsilon j} 
B^{\varepsilon j}_{\gamma}
+ g_2 \bar B^{\delta}_{\varepsilon i} q^{\gamma i}_{\delta j} 
B^{\varepsilon j}_{\gamma}
\end{equation}
Here greek indices run from 1 to 3 and latin indices from 1 to 2. For both
types of indices the Einstein summation convention is adopted. The 24
$q$-variables represent the pseudoscalar field in a P-state with respect to
the bare baryons. The 24 $p$-variables are the associated momenta. The $\bar
B$- and $B$-variables (16 of both) are the baryon creation and annihilation
operators. The variables are traceless:
\begin{equation}
q^{\alpha i}_{\beta i} = p^{\alpha i}_{\beta i} = 
q^{\alpha i}_{\alpha j} = p^{\alpha i}_{\alpha j} = 
B^{\alpha i}_{\alpha} = \bar B_{\alpha i}^{\alpha} = 0
\end{equation}
If the index K ($=1,\ldots,24$) distinguishes between the different field
and momentum variables the following relations are valid:
\begin{equation}
[p_{K},q_{K'}] = -i \delta_{KK'}
\end{equation}
Finally, if $|\rangle$ denotes the baryon vacuum we have:
\begin{equation}
B_{\delta}^{\gamma j}\bar B^{\mu}_{\nu k}\;|\rangle
=\left(\delta^\gamma_\nu\delta^\mu_\delta-
\frac{1}{3}\delta^\gamma_\delta\delta^\mu_\nu\right)\delta_k^j\;|\rangle
\end{equation}
the $\delta_j^i$ and the $\delta_\beta^\alpha$ being the Kronecker delta
symbols in 2 and 3 dimensions respectively.

\begin{figure}
\begin{center}
\hspace*{2cm}Relative level height\hspace{2em}Multiplet structure (SU(3), spin)\par
\vspace{2ex}
\includegraphics[width=4.5cm]{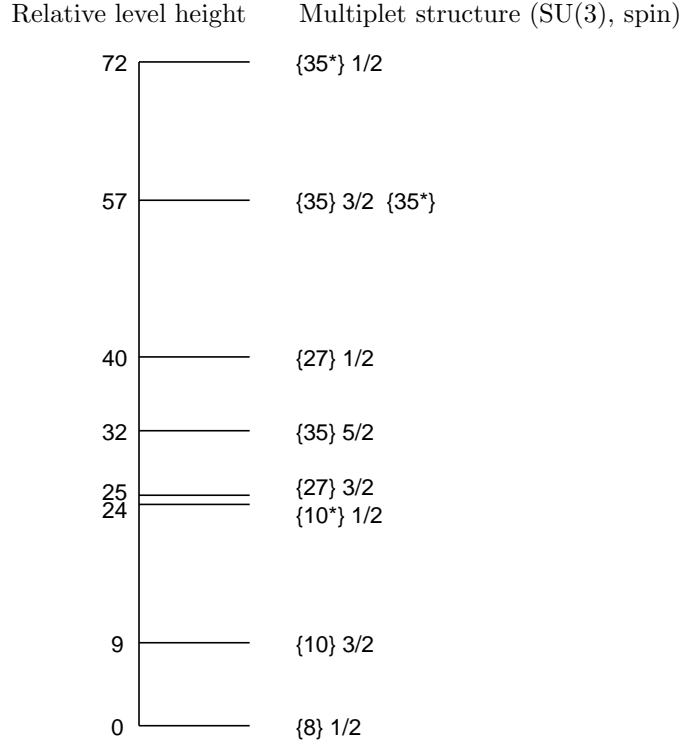}
\end{center}
\caption{Mass spectrum of baryon SU(3) multiplets (arbitrary units)}
\end{figure}

Eigenstates are sought of the form
\begin{equation}
\xi(q^{\alpha i}_{\beta j})\left(v_\nu^{\mu k}
\bar B^\nu_{\mu k}\right)\;|\rangle
\end{equation}
where $\xi$ has the index structure characteristic of the multiplet under
consideration.

In first instance, when $g_1$ and $g_2$ are large, the kinetic energy term
can be neglected. The first problem is then to find the value of $q^{\alpha
i}_{\beta j}$, with $q^2$ and $g_1/g_2$ constant, for which the operator
\begin{equation}
H' = \frac{1}{2}\mu^2 q^{\alpha i}_{\beta j} q_{\alpha i}^{\beta j}
+ g_1 \bar B^{\gamma}_{\delta i} q^{\delta i}_{\varepsilon j} 
B^{\varepsilon j}_{\gamma}
+ g_2 \bar B^{\delta}_{\varepsilon i} q^{\gamma i}_{\delta j} 
B^{\varepsilon j}_{\gamma}
\end{equation}
has the lowest possible eigenvalue. If one writes $q^{\alpha i}_{\beta j}$
in the form of a 3$\times$8 matrix, written as the combination of a
3$\times$3 and a 3$\times$5 matrix, then the minimum occurs when the first
matrix is proportional to the identity and the second matrix is zero. This
is the standard form. Any other 3$\times$8 matrix for which this minimum
occurs can be obtained from it by applying symmetry transformations of both
kinds. The simplicity of the standard form and the existence of symmetry
transformations leaving this form invariant guarantee that the final
spectrum of energy eigenstates is simple.  To find this spectrum the kinetic
term in the hamiltonian, previously omitted, must be taken into
account. Like in ordinary quantum mechanics, it must be split into a
``radial'' part and an ``angular'' part. However, there are now 13 ``radius-like''
variables and 11 ``angle-like'' variables. Still, the splitting can be carried
out. In the strong coupling limit the radial wave equations become
irrelevant. The differential equations associated with the angular part of
the hamiltonian lead to the desired spectrum. This is presented in the
figure.

The above method has been described in detail in reference [4]. An
alternative method leading to the same results has been presented by Goebel
[5].

\section*{References}

\noindent
$[1]$ S. Stepanyan et al., CLAS collaboration, 2003 hep-ex/0307018 ; C. Alt et al., 2003 hep-ex/0310014\\
$[2]$ W. Pauli and S.M. Dancoff, Phys. Rev. 62, 85 (1942); G. Wentzel, Rev. Mod. Phys. 19, 1 (1947), Phys. Rev. 125, 771 (1962) and Phys. Rev. 129, 1367 (1963) \\
$[3]$ C. Dullemond, Ann. Phys. (N.Y.) 33, 214 (1965); G. Wentzel, Suppl. Prog. Theoret. Phys. Commemoration Issue, 108 (1965)\\
$[4]$ F.J.M. von der Linden and C. Dullemond, Ann. Phys. (N.Y.) 41, 372 (1967)\\
$[5]$ C. Goebel, Phys. Rev. Lett. 16, 1130 (1966)

\end{document}